# On the Relationship Between Transit Time of ICMEs and Strength of the Initiated Geomagnetic Storms

I.M. Chertok[1]

**Abstract**  More than 140 isolated non-recurrent geomagnetic storms (GMSs) of various intensities from extreme to weak are considered, which are reliably identified with solar eruptive sources (coronal mass ejections, CMEs). The analysis aims to obtain a possibly complete picture of the relationship between the transit time of propagation of CMEs and interplanetary coronal mass ejections (ICMEs) from the Sun to the Earth (more precisely, the time interval $\Delta t_p$ from the moment of an eruption until the peak of the corresponding GMS) and the maximum intensity of this GMS, as measured by the disturbance storm time geomagnetic index Dst. Two groups of events are singled out: one includes GMSs, the source of which was an eruption from an active region (AR events), the other GMSs caused by filament eruptions from quiescent areas of the Sun located outside ARs (QS events). The distribution of the large number of the analyzed events on a $\Delta t_p$ – Dst plane confirms and substantially clarifies the known regularities. The AR events are characterized by a shorter transit time ($\Delta t_p \approx$ 1–4 days) and much stronger GMSs (Dst up to –600 nT mainly) in comparison with the QS events ($\Delta t_p \approx$ 3–5 days, Dst > –200 nT). For events of both groups, the shorter transit time of CMEs/ICMEs, the more intense GMSs; in particular, for AR events when $\Delta t_p$ declines from 4 to 1 day, Dst decreases on average from –100 to –470 nT and can reach – 900 nT. From the point of view of the nature of GMSs and their sources on the Sun, the obtained results mean that both the speed of CMEs/ICMEs and the strength of the magnetic field transferred by them are largely determined by the parameters of the corresponding eruptions, in particular, by the eruptive magnetic flux and the released energy.

## 1. Introduction

Among space weather disturbances that have a serious impact on the Earth's environment and modern technological systems, including satellites, power grids, communications, and navigation, solar energetic particle (SEP) events and geomagnetic storms (GMSs) are the most significant  (e.g. reviews by Gopalswamy, Tsurutani, and Yan, 2015; Kilpua, Koskinen , and Pulkkinen, 2017; Gopalswamy, 2018).  SEPs are accelerated in solar flares and coronal shocks driven by high-speed coronal mass ejections (CMEs), observed in the solar corona with the white-light ground-based and space-based coronagraphs. Relatively weak recurrent GMSs are related to so-called corotating interaction regions (SIRs) formed when high-speed streams of solar wind from coronal holes collide with the slower background wind.

----------------------------------------------------------------------
Corresponding author: I.M. Chertok
  ichertok@izmiran.ru

[1] Pushkov Institute of Terrestrial Magnetism, Ionosphere and Radio Wave Propagation (IZMIRAN), Troitsk, Moscow 108840, Russia

ORCID
I.M. Chertok    https://orcid.org/0000-0002-6013-5922

In this article, we deal with the most significant non-recurrent GMSs which are initiated by CMEs and their interplanetary counterparts (ICMEs). Most of such GMSs are caused by the front-sided halo CMEs originated from significant eruptions in the central sector of the visible solar disk (Gosling, 1993; Webb and Howard 2012; Gopalswamy, 2016; Ameri and Valtonen, 2017, and references therein). Two types of eruptions, associated CMEs/ICMEs, and resulting non-recurrent GMSs should be distinguished. The largest of them are usually associated with powerful flares and (when a filament is present) concomitant filament eruptions occurring in active regions (ARs) with their strong, complicated, and evolving magnetic fields. We will refer to this type of events as AR-associated or simply AR events. Noticeable but less intense GMSs are generated when CMEs occur as a result of erupting or disappearing solar filaments (DSFs) located in quiescent regions outside ARs. Following Zhang *et al*. (2007a), the non-AR events of this type we will mark as QS, i.e. quiet-Sun filament eruptions. For both AR and QS eruptions two factors are decisive for the occurrence of a significant GMS: (a) a perfect hit of ICMEs (magnetic clouds or general ejecta) on the Earth's magnetosphere, (b) the presence of a sufficiently strong and prolonged south-oriented (negative) $B_z$ component of the magnetic field, either in a ICME-driven shock and sheath region or in the ejecta body itself (see Manchester *et al*., 2017; Gopalswamy, 2018).

From the scientific and forecasting points of view, the most important parameters of non-recurrent GMSs are the maximum strength of the storm and its onset and peak times. In general, it is known that stronger GMSs occur earlier. This tendency follows, in particular, from the results of a number of studies of some restricted sets of events that show that the GMS intensity enhances with an increase in the initial speed of the corresponding CMEs in the inner corona and with a decrease of the ICME transit time from the Sun to the Earth (e.g. Srivastava and Venkatakrishnan, 2004; Lefèvre *et al*., 2016; Dumbović *et al*., 2016). Herewith, the time interval between a solar eruption and GMS onset manifested by a sudden storm commencement (SSC), $\Delta t_0$, was mainly used as the transit time. It was noted additionally that GMSs associated with QS eruptions are characterized as a whole by a weaker strength and larger time delay of their GMS onset in comparison with GMSs initiated by AR eruptions (Cliver *et al*., 2009; Dave *et al*., 2018). Moreover, it is known that the most severe GMSs, including the largest Carrington event of September 1859 and other historical superstorms, had the shortest time delays to the solar eruptive flares, being it of up to 17 hours (Cliver and Svalgaard, 2004; Cliver and Dietrich, 2013). It may be added that Chertok *et al*. (2013) determined reversed dependencies of the GMS strength and transit time on the CME-caused eruptive magnetic flux of extreme ultraviolet dimmings and post-eruption arcades and, on this basis, also concluded that short transit times are typical of the most intense GMSs.

This article aims to study the relationship between the GMS strength and the solar-Earth transit time in a more detailed and complete way using the entire sample of accessible data on GMSs of different strengths and their solar sources, from extreme events of the Carrington type to moderate and weak GMSs. Overall, our analysis covers the period from 1859 to 2018. We will characterize the GMS strength by the disturbance storm time, Dst, geomagnetic index (Sugiura and Kamei, 1991). For historical events, the minimum Dst value was estimated by geomagnetic magnetometer records available at that time and by the low-latitude boundary of the auroral oval (see Cliver and Svalgaard, 2004; Hapgood, 2019). Since 1957, the Dst index has been compiled regularly. Unlike previous studies, instead of the $\Delta t_0$ onset transit time, we will consider an interval between the flare maximum in the optics and soft X-rays (or the time of the CME eruption) and the moment of the GMS peak (the minimum of the Dst index), i.e. the peak transit time, $\Delta t_p$. The large sample of the analyzed events is described in Section 2. In Section 3, the generalized relationships between $\Delta t_p$ and Dst are presented both for AR and QS events. Section 4 is devoted to the summary and discussion.

## 2. Data

We consider isolated (i.e. undistorted by previous events) storms for which the estimated or measured Dst index is known and which are reliably enough, or with high probability, identified with a concrete solar eruptive source. The corresponding event number, date and time of the GMS peak, minimum Dxt index value, day, time, and type of a parent eruption, calculated transit peak time, as well as notes and references are presented in Table 1. We limited ourselves only to these parameters and do not provide other supporting information, in particular, such as the coordinates of eruptive flares and filaments, since the bulk of the events were related to sources in the central sector of the disk restricted by heliolongitudes ±45°. This information can be found in the given references. Some rare exceptions to this rule are noted by a superscript "c" in Table 1.

We formed our ensemble of events from different available sources. In order to consider a wide range of the GMS intensities, we included not only extreme events of the space era that are well provided with observational data, but also historical superstorms. A collection of such superstorms with Dst <–250 nT for 1859–2004 and their solar sources was listed and commented by Cliver and Crooker (1993), Cliver and Svalgaard (2004), Cliver *et al*. (2009) (see also references therein). Additionally, we checked the selected events with the solar data related to historical extreme GMSs in the range 1868–2010 and the space era presented by Gonzalez *et al*. (2011) and Lefèvre *et al*. (2016) (see also Joselyn and McIntosh, 1981, Tsurutani *et al*., 1992). Our analysis is based also on the catalog of major GMSs prepared by the Living with a Star (LWS) Coordinated Data Analysis Workshop (CDAW; Zhang *et al*., 2007a,b), that contains data on the most intense GMSs with a minimum Dst < −100 nT and their solar sources occurred during Solar Cycle 23 (1996–2005). We took into account some corrections to this catalog made by Chertok *et al*. (2013). The weak Cycle 24 was accompanied by a small number of major GMSs with Dst < −100 nT, a list of which up to June 2014 can be found, for example, in Gopalswamy, Tsurutani, and Yan (2015). Later, the sample of such events was supplemented by other eight major GMSs listed at the bottom of Table 1, the last of which, but the third largest GMS of Cycle 24, took place near its minimum on 26 August 2018 (see Li, Luhmann, and Lynch, 2018; Abunin *et al*., 2020; Chen *et al*., 2019 ). Concerning low-intensity events, we found and included in the analysis about 35 randomly selected isolated and moderate (−50 ≤ Dst < −100 nT) GMSs with a sufficiently reliable identification of the solar source. For this, we analyzed solar activity using the NOAA Solar-Geophisical Data (ftp://ftp.ngdc.noaa.gov/STP/SOLAR_DATA/SGD_PDFversion) and took into account the catalog of CMEs from the Large Angle and Spectrographic Coronagraph (LASCO) onboard the Solar and Heliospheric Observatory (SOHO) (i.e. the SOHO/LASCO CME catalog) of the CDAW Data Center at Goddard Space Flight Center compiled and maintained by NASA and the Catholic University of America in cooperation with the Naval Research Laboratory (https://cdaw.gsfc.nasa.gov/CME_list/; Yashiro *et al*., 2004; Gopalswamy *et al*., 2009) as well as the list of near-Earth ICMEs (http://www.srl.caltech.edu/ACE/ASC/DATA/level3/icmetable2.htm) collected and maintained by Richardson and Cane (2010).

When calculating the peak transit time, $\Delta t_p$, the eruption moment is well determined with an accuracy of ±1 hr. If there were no CME observations, in particular, for historical events, the time of maximum of optical or soft X-ray flares was taken as the eruption time. When the CME data were available, the information on the first appearance of a CME and its height-time trajectory, presented in the SOHO/LASCO CME catalog cited above, was used.

As a measure of the GMS intensity, the Dst index is used throughout this article, which started to be constructed and published only since 1957 (Sugiura and Kamei, 1991). Recall that the hourly Dst index is calculated from measurements of the H component of the field (the horizontal intensity of the magnetic field vector) at four low-latitude geomagnetic observatories and characterizes the effect of the magnetospheric global equatorial ring current. We proceeded from the final (1957–2014), provisional (2015–2016), and real-time values of Dst index presented at the World Data Center for Geomagnetism, Kyoto (http://wdc.kugi.kyoto-u.ac.jp/Dstdir/index.html). For earlier and historical events we use the Dst index estimated by different authors, based on available magnetogram records, worldwide low-latitude auroral extent, and other relevant data (see Cliver and Dietrich (2013) for details).

Among historical superintense and extreme GMSs, the type of a solar source (AR or QS) is reliably known for a small number of events, particularly from the studies of Joselyn and McIntosh (1981), Cliver and Crooker (1993), Cliver *et al*. (2009), and others. For some of those events, their relation either to AR-associated flares or to eruptive filaments outside ARs remains uncertain. One of the reasons is the lack of complete observations of the Sun in the pre-space era. Another category of uncertain events are the so-called intermediate filament eruptions (e.g. Zou *et al*., 2019). In such eruptions, driving usually quite high-speed CMEs, one filament leg is rooted in AR and another leg is located in a nearby quiet region, or lie between neighboring ARs. For both types of intense GMSs with uncertain and intermediate solar source association, we accepted that they were caused by AR eruptions as the most plausible origin.

Some remarks should be made regarding several of the analyzed events. Initial estimates indicated an extra large strength of the Carrington GMS of 1859 (event 1 in Table 1) up to Dst $\approx$ –1760 nT. However, a more detailed and complete analysis of various data and lines of evidence, including a probable ionospheric contribution, reviewed by Cliver and Dietrich (2013) resulted also in an extreme but more modest value Dst $\approx$ –900 nT. We use just this GMS intensity and the peak transit time $\Delta t_p \approx 18$ hr determined using the presented H-component magnetometer record.

The extreme 15 May 1921 GMS (event 4) can be regarded as a rival of the Carrington event. Hapgood (2019) reviewed available solar and geophysical data and reconstructed the timeline and scale of the May 1921 event. Following his work, we accepted that this superstorm was initiated by the flare eruption from a large near-central AR on 14 May (SOL1921-05-14T07) and the storm intensity also attained Dst $\approx$ –900 nT.

Vaisberg and Zastenker (1976) determined the optical 3B-class flare of 4 August 1972 (SOL1972-08-04T06:40) as the source of the moderate GMS (Dst $\approx$ –125 nT) that peaked on 5 August (event 25) and assigned a fast Sun-Earth travel time of about 15 hr for the corresponding interplanetary shock. We believe that this is not the case because the most probable source of this GMS was the H-alpha 3N-class flare on 2 August (SOL1972-08-02T04:10) from the same central AR. The three combined flares of 4 August probably contributed to the subsequent moderate GMSs that occurred 5–6 August.

The famous most powerful GMS of the space era, that picked on 14 March 1989 (event 34), has been associated usually with a series of solar eruptions from a large central AR between 10 and 12 March. According to Gopalsvamy *et al*. (2018) and Boteler (2019), the peak of this storm is attributed to the M7.3-class flare, which occurred on 12 March (SOL1989-03-12T00:16). Therefore, if so, its peak transit time is $\Delta t_p \approx 50$ hr that is still quite large for a storm of Dst $\approx$ –589 nT.

**Table 1** Analyzed geomagnetic storms, their solar sources, and the ICME transit time

| No. | Geomagnetic storm | | | Solar source | | $\Delta t_p$ (hr) | References |
|---|---|---|---|---|---|---|---|
| | Date (YYYY-MM-DD) | Peak time (hr) | Dst (nT) | Date, time (MM-DD, hr) | Type | | |
| 1 | 1859-09-02 | 06 | −900[a] | 09-01, 12 | AR | 18 | 1 |
| 2 | 1882-11-17 | 09 | −386[a] | 11-16, 06 | AR | 27 | 2 |
| 3 | 1909-09-25 | 12 | −595[a] | 09-24, 10 | AR | 26 | 3–5 |
| 4 | 1921-05-15 | 06 | −900[a] | 05-14, 07 | AR | 23 | 1,6 |
| 5 | 1938-01-22 | 11 | −344[a] | 01-20, 18 | AR | 41 | 2,7 |
| 6 | 1938-01-25 | 23 | −352[a] | 01-24, 14[c] | AR | 33 | 8 |
| 7 | 1940-03-24 | 20 | −366[a] | 03-23, 11 | AR | 33 | 7,9 |
| 8 | 1941-03-01 | 18 | −382[a] | 02-27, 20 | AR | 46 | 2,7 |
| 9 | 1941-07-05 | 13 | −453[a] | 07-03, 15 | AR | 46 | 7,9 |
| 10 | 1941-09-19 | 06 | −359[a] | 09-17, 09 | AR | 45 | 3,7 |
| 11 | 1946-03-28 | 14 | −440[a] | 03-27, 04 | AR | 34 | 2,7 |
| 12 | 1949-01-25 | 24 | −350[a] | 01-23, 02 | AR | 46 | 2,7 |
| 13 | 1957-01-21 | 23 | −250 | 01-20, 11 | AR | 36 | 10 |
| 14 | 1957-03-02 | 08 | −255 | 02-28, 00 | AR | 56 | 10 |
| 15 | 1957-09-13 | 11 | −427 | 09-11, 02 | AR | 57 | 10 |
| 16 | 1958-02-11 | 12 | −426 | 02-09, 22 | AR | 38 | 10 |
| 17 | 1958-07-08 | 21 | −327 | 07-07, 00 | AR | 45 | 10 |
| 18 | 1958-09-04 | 23 | −302 | 08-31, 01 | QS | 95 | 10 |
| 19 | 1959-07-15 | 20 | −429 | 07-14, 04 | AR | 40 | 10 |
| 20 | 1960-04-30 | 19 | −325 | 04-29, 01 | AR | 42 | 10 |
| 21 | 1960-11-13 | 10 | −339 | 11-11, 04 | AR | 54 | 10 |
| 22 | 1961-10-28 | 19 | −272 | 10-25-26[b] | QS | 67 | 10 |
| 23 | 1967-05-26 | 06 | −387 | 05-23, 18 | AR | 60 | 10 |
| 24 | 1970-03-08 | 23 | −284 | 03-07, 01 | AR | 46 | 10 |
| 25 | 1972-08-05 | 04 | −125 | 08-02, 04 | AR | 72 | 11 |
| 26 | 1977-08-05 | 09 | −61 | 08-01[b] | QS | 93 | 12 |
| 27 | 1978-07-05 | 07 | −99 | 06-30[b] | QS | 115 | 12 |
| 28 | 1978-08-28 | 09 | −222 | 08-22-23[b] | QS | 129 | 12 |
| 29 | 1979-01-07 | 21 | −100 | 01-03[b] | QS | 93 | 12 |
| 30 | 1979-04-25 | 15 | −149 | 04-19-20[b] | QS | 135 | 12 |
| 31 | 1980-12-19 | 19 | −240 | 12-16, 12 | AR | 79 | 13 |
| 32 | 1982-07-14 | 02 | −325 | 07-12, 09 | AR | 41 | 7,10,13 |
| 33 | 1982-09-06 | 12 | −289 | 09-04, 00 | AR | 60 | 10,13 |
| 34 | 1989-03-14 | 02 | −589 | 03-12, 00 | AR | 50 | 14,15 |
| 35 | 1989-09-19 | 05 | −255 | 09-15, 22 | AR | 79 | 10 |
| 36 | 1989-11-17 | 23 | −266 | 11-15, 06 | AR | 65 | 10 |
| 37 | 1990-04-10 | 19 | −281 | 04-08, 03 | AR | 64 | 10 |
| 38 | 1991-11-09 | 02 | −354 | 11-05, 22 | QS | 76 | 7,8 |
| 39 | 1994-04-17 | 08 | −201 | 04-14 04 | QS | 76 | 16 |
| 40 | 1997-01-10 | 10 | −78 | 01-06, 15 | QS | 91 | 17 |
| 41 | 1997-04-11 | 05 | −82 | 04-07, 14 | AR | 87 | 17 |
| 42 | 1997-05-15 | 13 | −115 | 05-12, 05 | AR | 80 | 18 |
| 43 | 1997-09-03 | 23 | −98 | 08-29, 24 | AR | 95 | 17 |
| 44 | 1997-10-01 | 16 | −98 | 09-28, 01 | QS | 87 | 17 |
| 45 | 1997-10-11 | 04 | −130 | 10-06, 15 | QS | 109 | 18 |
| 46 | 1997-11-07 | 05 | −110 | 11-04, 06 | AR | 71 | 18 |
| 47 | 1997-12-30 | 20 | −77 | 12-26, 02[c] | QS | 114 | 17 |
| 48 | 1998-01-07 | 05 | −77 | 01-02, 23 | QS | 102 | 17 |
| 49 | 1998-02-18 | 01 | −100 | 02-14, 05 | QS | 92 | 17 |
| 50 | 1998-05-02 | 18 | −85 | 04-29, 16 | AR | 74 | 17 |
| 51 | 1998-05-04 | 06 | −205 | 05-02, 14 | AR | 40 | 18 |
| 52 | 1998-06-26 | 05 | −101 | 06-22, 07 | QS | 94 | 18 |
| 53 | 1998-08-27 | 10 | −155 | 08-24, 22 | AR | 60 | 18 |
| 54 | 1998-09-25 | 10 | −207 | 09-23, 07 | AR | 51 | 18 |
| 55 | 1998-10-19 | 16 | −112 | 10-15, 10 | QS | 102 | 18 |
| 56 | 1998-11-13 | 22 | −131 | 11-09, 18 | QS | 100 | 18 |
| 57 | 1999-02-18 | 10 | −123 | 02-16, 03 | AR | 55 | 18 |

**Table 1** (Continued)

| No. | Geomagnetic storm | | | Solar source | | $\Delta t_p$ (hr) | References |
|---|---|---|---|---|---|---|---|
| | Date (YYYY-MM-DD) | Peak time (hr) | Dst (nT) | Date, time (MM-DD, hr) | Type | | |
| 58 | 1999-03-01 | 01 | –94 | 02-24, 16 | QS | 107 | 11 |
| 59 | 1999-04-17 | 08 | –91 | 04-13, 02 | QS | 102 | 17 |
| 60 | 1999-09-22 | 24 | –173 | 09-20, 06 | QS | 66 | 18 |
| 61 | 1999-10-22 | 07 | –237 | 10-17, 23 | QS | 104 | 18 |
| 62 | 2000-01-23 | 01 | –97 | 01-18, 17 | AR | 104 | 17 |
| 63 | 2000-02-12 | 12 | –133 | 02-09, 20 | AR | 64 | 18,19 |
| 64 | 2000-04-07 | 01 | –288 | 04-04, 16[c] | AR | 57 | 18 |
| 65 | 2000-05-17 | 06 | –91 | 05-13, 12 | AR | 90 | 17 |
| 66 | 2000-05-24 | 09 | –147 | 05-21, 07 | AR | 74 | 17 |
| 67 | 2000-06-08 | 20 | –90 | 06-06, 15 | AR | 53 | 17 |
| 68 | 2000-07-16 | 01 | –301 | 07-14, 11 | AR | 38 | 18 |
| 69 | 2000-08-11 | 07 | –105 | 08-06, 22 | AR | 105 | 18 |
| 70 | 2000-08-12 | 10 | –235 | 08-09, 16 | AR | 66 | 18 |
| 71 | 2000-10-29 | 04 | –127 | 10-25, 08[c] | QS | 92 | 18 |
| 72 | 2000-11-06 | 22 | –159 | 11-03, 18 | AR | 76 | 18,19 |
| 73 | 2000-11-27 | 02 | –79 | 11-24, 15 | AR | 59 | 20 |
| 74 | 2000-12-23 | 05 | –62 | 12-18, 11 | AR | 114 | 20 |
| 75 | 2001-03-05 | 03 | –73 | 02-28, 14 | QS | 129 | 17 |
| 76 | 2001-03-31 | 09 | –387 | 03-29, 10 | AR | 47 | 18 |
| 77 | 2001-04-11 | 24 | –271 | 04-10, 05 | AR | 43 | 18 |
| 78 | 2001-08-17 | 22 | –105 | 08-14, 14 | QS | 80 | 18 |
| 79 | 2001-09-26 | 02 | –102 | 09-24, 10 | AR | 40 | 18 |
| 80 | 2001-10-21 | 22 | –187 | 10-19, 16 | AR | 54 | 18 |
| 81 | 2001-10-28 | 12 | –157 | 10-25, 15 | AR | 69 | 18 |
| 82 | 2001-11-06 | 07 | –292 | 11-04, 16 | AR | 39 | 18 |
| 83 | 2001-11-24 | 17 | –221 | 11-22, 23 | AR | 42 | 18 |
| 84 | 2002-02-02 | 10 | –86 | 01-28, 10 | QS | 120 | 11 |
| 85 | 2002-05-11 | 20 | –110 | 05-08, 14 | AR | 78 | 18 |
| 86 | 2002-08-02 | 06 | –102 | 07-29, 12 | QS | 90 | 18 |
| 87 | 2002-09-08 | 01 | –181 | 09-05, 17 | AR | 55 | 18 |
| 88 | 2003-08-18 | 16 | –148 | 08-14, 19 | AR | 93 | 18,19 |
| 89 | 2003-10-30 | 01 | –353 | 10-28, 11 | AR | 38 | 18 |
| 90 | 2003-10-30 | 23 | –383 | 10-29, 21 | AR | 26 | 18 |
| 91 | 2003-11-20 | 21 | –422 | 11-18, 09 | AR | 60 | 18,21 |
| 92 | 2004-01-22 | 14 | –130 | 01-20, 01 | AR | 61 | 18 |
| 93 | 2004-04-04 | 01 | –117 | 03-31, 11 | AR | 86 | 18 |
| 94 | 2004-07-23 | 03 | –99 | 07-20, 12 | AR | 63 | 18 |
| 95 | 2004-07-25 | 17 | –136 | 07-22, 08 | AR | 81 | 18 |
| 96 | 2004-07-27 | 14 | –170 | 07-25, 14 | AR | 48 | 18 |
| 97 | 2004-08-30 | 23 | –129 | 08-26, 12 | QS | 107 | 18,19 |
| 98 | 2004-11-08 | 07 | –374 | 11-06, 00 | AR | 55 | 18,19 |
| 99 | 2005-01-08 | 03 | –93 | 01-05, 13 | QS | 62 | 17 |
| 100 | 2005-01-22 | 06 | –97 | 01-20, 07[c] | AR | 47 | 18 |
| 101 | 2005-05-15 | 09 | –247 | 05-13, 17 | AR | 40 | 18 |
| 102 | 2005-05-20 | 09 | –83 | 05-16, 13 | AR | 89 | 18 |
| 103 | 2005-06-13 | 01 | –106 | 06-09, 14 | AR | 83 | 18 |
| 104 | 2005-08-24 | 12 | –184 | 08-22, 01[c] | AR | 59 | 18 |
| 105 | 2006-08-20 | 02 | –79 | 08-16, 16 | AR | 82 | 17 |
| 106 | 2006-12-15 | 08 | –162 | 12-13, 02 | AR | 54 | 17 |
| 107 | 2010-04-06 | 15 | –81 | 04-03, 10 | AR | 67 | 20 |
| 108 | 2010-04-13 | 02 | –67 | 04-08, 04 | AR | 94 | 22 |
| 109 | 2010-05-29 | 13 | –80 | 05-23, 18 | QS | 139 | 23 |
| 110 | 2010-08-04 | 02 | –74 | 08-01, 08 | QS | 66 | 17 |
| 111 | 2010-10-11 | 20 | –75 | 10-06, 07 | QS | 133 | 23 |
| 112 | 2011-08-06 | 04 | –115 | 08-04, 04 | AR | 48 | 17 |
| 113 | 2011-09-17 | 16 | –72 | 09-13, 22 | AR | 90 | 17 |

**Table 1** (Continued)

| No. | Geomagnetic storm Date (YYYY-MM-DD) | Peak time (hr) | Dst (nT) | Solar source Date, time (MM-DD, hr) | Type | $\Delta t_p$ (hr) | References |
|---|---|---|---|---|---|---|---|
| 114 | 2011-09-26 | 24 | –118 | 09-24, 12[c] | AR | 60 | 24 |
| 115 | 2011-10-25 | 07 | –134 | 10-22, 00 | QS | 79 | 24 |
| 116 | 2012-01-22 | 22 | –70 | 01-19, 14 | AR | 80 | 17 |
| 117 | 2012-03-09 | 09 | –145 | 03-07, 00 | AR | 57 | 24 |
| 118 | 2012-04-24 | 05 | –120 | 04-19, 15[c] | QS | 110 | 24 |
| 119 | 2012-06-17 | 14 | –86 | 06-14, 14 | AR | 72 | 20 |
| 120 | 2012-07-15 | 19 | –139 | 07-12, 16 | AR | 75 | 24 |
| 121 | 2012-07-24 | 05 | –900[a] | 07-23, 02[d] | AR | 27 | 25,26 |
| 122 | 2012-10-01 | 05 | –122 | 09-28, 00 | AR | 77 | 24 |
| 123 | 2012-11-14 | 08 | –108 | 11-09, 15 | QS | 113 | 24 |
| 124 | 2013-03-17 | 21 | –132 | 03-15, 07 | AR | 62 | 24 |
| 125 | 2013-07-14 | 23 | –81 | 07-09, 15 | QS | 128 | 17 |
| 126 | 2013-10-02 | 08 | –72 | 09-29, 22 | QS | 58 | 17 |
| 127 | 2014-02-19 | 09 | –119 | 02-16, 09 | AR | 72 | 24 |
| 128 | 2014-09-12 | 24 | –88 | 09-10, 17 | AR | 56 | 17 |
| 129 | 2015-03-17 | 23 | –222 | 03-15, 02 | QS | 69 | 27 |
| 130 | 2015-06-23 | 05 | –204 | 06-21, 02 | AR | 51 | 27 |
| 131 | 2015-08-16 | 08 | –84 | 08-12, 14 | QS | 90 | 17 |
| 132 | 2015-09-09 | 13 | –98 | 09-04, 17 | QS | 116 | 17 |
| 133 | 2015-11-07 | 07 | –89 | 11-04,13 | AR | 66 | 28 |
| 134 | 2015-12-20 | 23 | –155 | 12-16, 09 | QS | 110 | 17 |
| 135 | 2016-01-01 | 01 | –110 | 12-28, 12 | AR | 85 | 17 |
| 136 | 2016-01-20 | 17 | –93 | 01-14, 20 | QS | 141 | 17 |
| 137 | 2016-03-05 | 22 | –98 | 03-03, 15 | QS | 79 | 11 |
| 138 | 2016-10-13 | 18 | –103 | 10-08, 15 | QS | 123 | 17,29 |
| 139 | 2017-05-28 | 08 | –125 | 05-23, 05 | QS | 123 | 17 |
| 140 | 2017-07-16 | 16 | –72 | 07-14, 01 | AR | 63 | 17 |
| 141 | 2017-09-08 | 02 | –124 | 09-06, 12 | AR | 38 | 17,30 |
| 142 | 2018-08-26 | 07 | –174 | 08-20, 21 | QS | 130 | 31,32 |
| 143[e] | 1903-10-31 | 14 | –513[a] | 10-30, 06 | AR | 32 | 33 |

[a]Esimated Dst value.
[b]Eruption time is known to within 1 day.
[c]Heliolongitude of the eruption source exceeds 45° from the central meridian.
[d]The event associated with a far behind the limb eruption observed by the favorably located STEREO-A spacecraft.
[e]The event is added at the last moment.

References: (1) Cliver, Dietrich (2013); (2) Lefèvre *et al.* (2016); (3) Newton (1943); (4) Love, Hayakawa, Cliver (2019); (5) Hayakawa *et al.* (2019); (6) Hapgood (2019); (7) Cliver, Svalgaard (2004); (8) Cliver *et al.* (2009); (9) Newton (1944); (10) Cliver, Crooker (1993); (11) This article; (12) Joselyn, McIntosh (1981); (13) Tsurutani *et al.* (1992); (14) Gopalswamy (2018); (15) Boteler (2019); (16) McAllister *et al.* (1996); (17) Richardson, Cane (2010); (18) Zhang *et al.* (2007a,b); (19) Chertok *et al.* (2013); (20) Ameri, Valtonen (2017); (21) Grechnev *et al.* (2014); (22) Davis *et al.* (2011); (23) Dave *et al.* (2018); (24) Gopalswamy, Tsurutani, Yan (2015); (25) Liu *et al.* (2014); (26) Baker *et al.* (2013); (27) Webb, Nitta (2017); (28) Wang *et al.* (2018); (29) He *et al.* (2018); (30) Wu *et al.* (2019); (31) Chen *et al.*(2019); (32) Abunin *et al.* (2020); (33) Hayakawa *et al.* (2020).

As an exception, we included in our consideration the outstanding event 121 associated with a far behind the limb eruption on 23 July 2012 (SOL2012-07-23T02:36), which was well observed by the favorably located STEREO-A (Solar Terrestrial Relations Observatory) spacecraft. The initiated ICME arrived at STEREO-A in a very short transit time of 19–21 hr and brought to 1 AU a recorded strong interplanetary magnetic field of 109 nT with a sustained southward $B_z$ component. According to modeling estimations by Liu *et al.* (2014) (see also Baker *et al.*, 2013), if it had been directed toward Earth, this ICME might have caused a GMS with a minimum Dst index between −1150 and −600 nT that is comparable to the Carrington storm of 1859. We took the average Dst ≈ –900 nT and the peak transit time $\Delta t_p$ ≈ 27 hr.

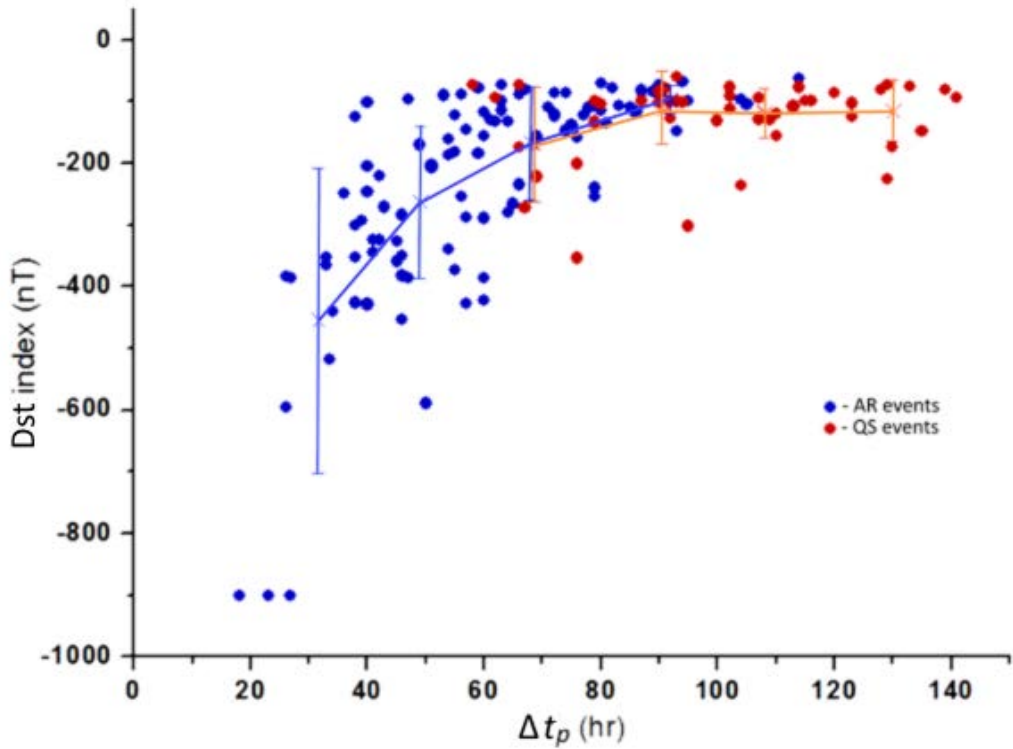

**Figure 1** Correlation between the minimal Dst index and peak transit time, $\Delta t_p$, values for the set of events listed in Table 1. Red circles represent events associated with AR eruptions and blue circles refer to QS filament eruptions. Crosses mark the corresponding bin-averaged values of the Dst index with standard statistical deviations.

A series of events was not included in our consideration since the corresponding GMSs occurred against a strongly disturbed and intense background from a previous storm and their real strength was not determined unambiguously by a relevant solar eruption. Examples of such overlapping events are the fairly intense GMSs of 5 and 23 September 1957 (Dst ≈ = –324 and 303 nT), 1 April and 7 October 1960 (Dst ≈ = –327 and –287 nT), 13 April 1981 (Dst ≈ = –311 nT), 9 February 1986 (Dst ≈ = –307 nT), and others.

The preprint of Hayakawa *et al*. (2020) has just been published with an analysis of the extreme GMS on 31 October 1903. We added it at the end of Table1 as event 143 and we included it in our study.

## 3. Results

The distribution of the minimal geomagnetic Dst index versus the peak transit time, $\Delta t_p$, for the analyzed large collection of events is presented in Figure 1. This plot confirms and clarifies the basic regularities mentioned in the Introduction. At first, the QS-associated GMSs (red circles) as a whole are characterized by the longer peak transit time and lower intensity in comparison with the AR-driven storms (blue circles). A noticeable part of QS GSMs peaked only over 5 days after a filament eruption, but only in several AR-associated storms the time interval between an eruption and storm maximum did not exceeded 90 hr. More than a half of the QS events have $\Delta t_p$ > 80 hr and Dst > –150 nT while the majority of the AR events have $\Delta t_p$ < 70 hr and Dst < –150 nT. The strength of the most intense QS-associated GMS of 9 November 1991 was Dst ≈ – 354 nT and the peak transit time $\Delta t_p$ ≈ 76 hr, and in the strongest AR storms of 2 September 1859, 15 May 1921, 23 July 2012 Dst reached –900 nT in the 18–27 hr after an eruption.

Secondly, in spite of the wide scatter, both the AR and QS events display an evident tendency that the storm strength increases with the decreasing peak transit time. In particular, this tendency is revealed by the strongest GMSs at the outer edge of the scatter plot of the whole sample of events. This is also evidenced by the averaged values of Dst calculated in several $\Delta t_p$ bins. It is clear that this tendency is less pronounced for GMSs associated with the QS filament eruptions due to a smaller range of the storm strength. For the QS events, the average GMS intensity remains almost unchanged (<Dst> ≈ –100 nT for the analyzed set of events) with a peak transient time in the range of $\Delta t_p$ > 80 hr and somewhat increases (to < Dst > ≈ –170 nT) for shorter $\Delta t_p$. For the AR events, the average storm strength monotonously and more and more sharply becomes stronger from < Dst > ≈ –100 nT at $\Delta t_p$ > 80 hr to < Dst > ≈ –450 nT with $\Delta t_p$ < 40 hr.

The large scatter of events in Figure 1 it is due to many of obvious and important factors (e.g. reviews by Kilpua, Koskinen, and Pulkkinen, 2017; Manchester *et al*., 2017; Yermolaev *et al*., 2018). In particular, the GMS intensity strongly depends on the ICME type (whether it is a magnetic cloud or an ejecta), on its internal structure including a sheath region and flux rope, on the magnetic field strength brought by an ICME to the Earth, on what part of its field falls on the negative $B_z$ component, on whether an ICME hits the magnetosphere with front or the glancing impact occurs, and others. In addition, it is known that the geoeffectiveness of solar eruptions is subjected to the pronounced seasonal variation, being larger around equinoxes and smaller at solstices (Cliver and Crooker, 1993). In turn, the ICME transit time depends on the initial velocity of a CME in the corona, on the ICME drag and deceleration by the background solar wind, on its interaction with various interplanetary structures, and so on. All these factors can match in a variety of combinations that leads to the observed scatter.

## 4. Summary and Discussion

We studied the large and updated dataset of non-recurrent GMSs with variegated intensity and their eruptive sources from the central sector of the solar disk. We considered about 140 isolated events over the period 1859–2018 and focused on the relationship between the ICME transit time and the GMS strength scaled by the Dst index. There are two additional differences from previous studies. For example the ICME transit time, the interval $\Delta t_p$ between the time of the solar eruption and the GMS peak (the minimum value of Dst), but not the GMS onset ($\Delta T_o$), was taken. Two types of GMSs initiated by CMEs/ICMEs formed by the flare-associated eruptions in active regions (AR events) and by the filament eruptions from quiescent areas of the Sun located outside ARs (QS events) were considered in a single view, but were distinguished from each other. Such an analysis made it possible to obtain a more comprehensive and visually clear picture of the relationship between the ICME transit time and the GMS strength.

Despite significant scatter of the CME/ICME and GMS characteristics caused by many factors described in the Section 3, the performed analysis of the extensive sample of events (Figure 1) reveals regularities which confirm and substantially clarify known trends (see Kim *et al*., 2010; Cliver and Dietrich, 2013; Dumbović *et al*., 2015; Gopalswamy, Tsurutani, and Yan, 2015; Ibrahim *et al*., 2019; Zou *et al*., 2019). Firstly, the AR events are characterized by the shorter ICME peak transit time and stronger storm intensity than the QS associated GMSs. In our sample, the AR storms display $\Delta t_p$ mainly between 20–90 hr and Dst up to extreme –900 nT, while for the QS storms these parameters vary within 60–140 hr and Dst > –350 nT. It is obvious that these differences between the two types of GMSs are due to that the AR eruptions occur in strong sunspot magnetic fields and the QS eruptions are associated with the relatively weak magnetic fields of the quiescent areas. Secondly, the GMS strength increases (Dst goes down) with decreasing peak transit time $\Delta t_p$. It is demonstrated by the strongest GMSs at the outer edge of the scatter plot of the whole sample of events as well as by the averaged values of Dst calculated in several $\Delta t_p$ bins. This important feature is characteristic for both AR- and QS events, but for AR storms it is more clear. In particular, the average strength of the AR-associated GMSs monotonously and more and more sharply becomes stronger from $<Dst> \approx -100$ nT at $\Delta t_p > 80$ hr to $<Dst> \approx -450$ nT at $\Delta t_p < 40$ hr. Thus, the most intense and rapidly advancing GMSs are certainly initiated by AR eruptions, and the weakest and severely delayed GMSs are caused by QS filament eruptions outside ARs. Based on this, it is reasonable to expect that GMSs, associated with eruptions of the intermediate type in relation to AR and QS ones, fall in the middle of the scatter plot (Figure 1) where AR and QS events overlap.

The regularities indicated above can be interpreted as an evidence that the main ICME parameters on which the intensity of the corresponding GMSs depends, such as the ICME speed and strength of the inner magnetic field, are largely determined by the characteristics of the solar eruptions, particularly, the eruptive magnetic flux. This is consistent, for example, with the finding of Chertok *et al.* (2013) that the GMS strength enhances and the ICME transit time decreases with increasing eruptive magnetic flux, which can be estimated, e.g., through analysis of extreme ultraviolet dimmings (see Dissauer *et al.*, 2018), post-eruption arcades (Gopalswamy *et al.*, 2017), and flare ribbons (Kazachenko *et al.*, 2017). Further, Pal *et al.* (2018) established the dependence of the CME properties on characteristics of their solar source active region and associated flare reconnection flux that in turn determines the severity and onset time of the associated GMSs. Further, sufficiently good direct correlations between the near-Sun CME sky-plane speed and the strength of the associated GMSs, especially for the magnetic cloud ICMEs, have been reported by, e.g., Srivastava and Venkatakrishnan (2004), Kim *et al.* (2010), Gopalswamy (2010), and Shanmugaraju *et al.* (2015). It is worth adding that, according to Michalek, Gopalswamy, and Yashiro (2008), the transit time to Earth of ICMEs, initiated by halo CMEs, reveals a clear reversed dependence on the deprojected space speed of the latter (see also Ibrahim *et al.*, 2019). This means that as a result of solar eruptions the highest speed ICMEs acquire such large kinetic energy that effectively overcomes the aerodynamic drag while they propagate in the solar wind. At the same time, such high-speed ICMEs bring to Earth sufficiently strong magnetic fields, that allows them, under favorable conditions, to cause severe and extreme GMSs. The comprehensive picture of the relationships between the travel time of ICMEs and the intensity of the corresponding GMSs presented in this article may be useful for further studies of the solar eruptions and predictions of their space weather impacts.

**Acknowledgements** The author thanks an anonymous reviewer for useful remarks and comments. The author is grateful to the teams and numerous colleagues whose data and results are used in this analysis. This research was partially supported by the Russian Foundation of Basic Research under grants 17-02-00308 and by the Complex Program 19–270 of the Russian Ministry of Education and Science.

# References

Abunin, A. A., Abunina, M. A., Belov, A. V., Chertok, I. M.: 2020, Peculiar solar sources and geospace disturbances on 20–26 August 2018. *Solar Phys*. **295**(1), 7. https://doi.org/10.1007/s11207-019-1574-8

Ameri, D., Valtonen, E.: 2017, Investigation of the geoeffectiveness of disk-centre full-halo coronal mass ejections. *Solar Phys*. **292**(6), 79. https://doi.org/10.1007/s11207-017-1102-7

Baker, D. N., Li, X., Pulkkinen, A., Ngwira, C. M., Mays, M. L., Galvin, A. B., Simunac, K. D. C.: 2013, A major solar eruptive event in July 2012: defining extreme space weather scenarios. *Space Weather*. **11**(10), 585–591. https://doi.org/10.1002/swe.20097

Boteler, D. H.: 2019, A twenty-first century view of the March 1989 magnetic storm. *Space Weather*. **17**(10), 1427–1441. https://doi.org/10.1029/2019SW002278

Chen, C., Liu, Y. D., Wang, R., Zhao, X., Hu, H., Zhu, B.: 2019, Characteristics of a gradual filament eruption and subsequent CME propagation in relation to a strong geomagnetic storm. *Astrophys. J.* **884**(1), 90. https://doi.org/10.3847/1538-4357/ab3f36

Chertok, I. M., Grechnev, V. V., Belov, A. V., Abunin, A. A.: 2013. Magnetic flux of EUV arcade and dimming regions as a relevant parameter for early diagnostics of solar eruptions - sources of non-recurrent geomagnetic storms and Forbush decreases. *Solar Phys.* **282(1)**, 175–199. https://doi.org/10.1007/s11207-012-0127-1

Cliver, E. W., Crooker, N. U.: 1993, A seasonal dependence for the geoeffectiveness of eruptive solar events. *Solar. Phys*., **145**(2), 347–357. https://doi.org/10.1007/BF00690661

Cliver, E. W., Dietrich, W. F.: 2013, The 1859 space weather event revisited: Limits of extreme activity. *J. Space Weather Space Clim*. **3**, A31. https://doi.org/10.1051/swsc/2013053

Cliver, E. W., L. Svalgaard: 2004, The 1859 solar-terrestrial disturbance and the current limits of extreme space weather activity. *Solar. Phys*. **224**(1–2), 407–422. https://doi.org/10.1007/s11207-005-4980-z

Cliver, E. W., Balasubramaniam, K. S., Nitta, N. V., Li, X.: 2009, Great geomagnetic storm of 9 November 1991: Association with a disappearing solar filament. *J. Geophys. Res*. **114**(3), A00A20. https://doi.org/10.1029/2008JA013232

Dave, K., Mishra W., Srivastava, N., Jadhav, R. M.: 2018, Study of interplanetary and geomagnetic response of filament associated CMEs. *Proceedings IAU Symposium*. No. 340, 83–84. https://doi.org/10.1017/S174392131800203X

Davis, C. J., de Koning, C. A., Davies, J. A., Biesecker, D., Millward, G., Dryer, M., *et al*.: 2011, A comparison of space weather analysis techniques used to predict the arrival of the Earth-directed CME and its shockwave launched on 8 April 2010. Space Weather. **9**(1), S01005. https://doi.org/10.1029/2010SW000620

Dissauer, K., Veronig, A. M., Temmer, M., Podladchikova, T., Vanninathan, K.: 2018, Statistics of coronal dimmings associated with coronal mass ejections. I. Characteristic dimming properties and flare association. *Astrophys. J.* **863**(2), 169. https://doi.org/10.3847/1538-4357/aad3c6

Dumbović, M., Devos, A., Vršnak, B., Sudar, D., Rodriguez, L., Ruždjak, D., *et al*.: 2015, Geoeffectiveness of coronal mass ejections in the SOHO era. *Solar Phys.* **290**(2), 579–612. https://doi.org/10.1007/s11207-014-0613-8

Gonzalez, W. D., Echer, E., de Gonzalez, A. L. C., Tsurutani, B. T., Lakhina, G. S.: 2011, Extreme geomagnetic storms, recent Gleissberg cycles and space era-superintense storms. *J. Atmos. Sol.–Terr. Phys.* **73**(11-12), 1447–1453. https://doi.org/10.1016/j.jastp.2010.07.023

Gopalswamy, N.: 2010, The CME link to geomagnetic storms. Proc. of the IAU Symp. **264**, 326–335. https://doi.org/10.1017/S1743921309992870

Gopalswamy, N.: 2016, History and development of coronal mass ejections as a key player in solar terrestrial relationship. *Geoscience Lett*. **3**, 8. https://doi.org/10.1186/s40562-016-0039-2

Gopalswamy, N.: 2018, Extreme solar eruptions and their space weather consequences. In: Extreme Events in Geospace. Origins, predictability, and consequences. Ed. by N. Buzulukova, Elsevier, p. 37–63. https://doi.org/10.1016/B978-0-12-812700-1.00002-9

Gopalswamy, N., Tsurutani, B., Yan, Y.: 2015, Short-term variability of the Sun-Earth system: an overview of progress made during the CAWSES-II period. *Progress in Earth and Planetary Science.* **2**, 13. https://doi.org/10.1186/s40645-015-0043-8

Gopalswamy, N., Yashiro, S., Akiyama, S., Xie, H.: 2017, Estimation of reconnection flux using post-eruption arcades and its relevance to magnetic clouds at 1 AU. *Solar Phys.* **292**(4), 65. https://doi.org/10.1007/s11207-017-1080-9

Gopalswamy, N., Yashiro, S., Michalek, G., Stenborg, G., Vourlidas, A., Freeland, S., Howard, R.: 2009, The SOHO/LASCO CME catalog. *Earth Moon Planets*, **104**(1–4), 295–313. https://doi.org/10.1007/s11038-008-9282-7

Gosling, J. T.: 1993, The solar flare myth. *J. Geophys. Res.* **98**(A11), 18,937–18, 949. https://doi.org/10.1029/93JA01896

Grechnev, V. V., Uralov, A. M., Chertok, I. M., Belov, A. V., Filippov, B. P., Slemzin, V. A., Jackson, B. V.: 2014, A challenging solar eruptive event of 18 November 2003 and the causes of the 20 November geomagnetic superstorm. IV. Unusual magnetic cloud and overall scenario. *Solar Phys.* **289**(12), 4653–4673. https://doi.org/10.1007/s11207-014-0596-5

Hapgood, M.: 2019, The great storm of May 1921: an exemplar of a dangerous space weather event. *Space Weather*, 17(7), 950–975. https://doi.org/10.1029/2019sw002195

Hayakawa, H., Ebihara, Y., Cliver, E. W., Hattori, K., Toriumi, S., Love, J. J., *et al.*: 2019, The extreme space weather event in September 1909. *Mon. Not. R. Astron. Soc.* **484**(3), 4083–4099. https://doi.org/10.1093/mnras/sty3196

Hayakawa, H., Ribeiro, P., Vaquero, J. M., Gallego, M. C., Knipp, D. J., Mekhaldi, F., *et al.*: 2020, The extreme space weather event in 1903 October/November: an outburst from the quiet Sun. *Astrophys. J. Lett*. https://arxiv.org/abs/2001.04575

He, W., Liu, Y. D., Hu, H., Wang, R., Zhao, X.; 2018, A stealth CME bracketed between slow and fast wind producing unexpected geo-effectiveness. *Astrophys. J.* **860**(1), 78. https://doi.org/10.3847/1538-4357/aac381

Ibrahim, M. S., Joshi, B., Cho, K.-S., Kim, R.-S., Moon, Y.-J.: 2019, Interplanetary coronal mass ejections during solar cycles 23 and 24: Sun–Earth propagation characteristics and consequences at the near-Earth region. *Solar Phys*. **294**(6), 54. https://doi.org/10.1007/s11207-019-1443-5

Joselyn, J. A., McIntosh, P. S.: 1981, Disappearing solar filaments: a useful predictor of geomagnetic activity. *J. Geophys. Res.* **86**(A6), 4555–4564. https://doi.org/doi:10.1029/ja086ia06p04555

Kazachenko, M. D., Lynch, B. J., Welsch, B., Sun, X.: 2017, A database of flare ribbon properties from the Solar Dynamics Observatory. I. Reconnection flux. *Astrophys. J*. **845**(1), 49. https://doi.org/10.3847/1538-4357/aa7ed6

Kim, R.-S., Cho, K.-S., Moon, Y.-J., Dryer, M., Lee, J., Yi, Y., *et al*.: 2010, An empirical model for prediction of geomagnetic storms using initially observed CME parameters at the Sun. *J. Geophys. Res*. **115**(A12), 12108. https://doi.org/10.1029/2010JA015322

Kilpua, E., Koskinen, H. E. J., Pulkkinen, T.: 2017, Coronal mass ejections and their sheath regions in interplanetary space. *Living Rev. Sol. Phys.* **14**, 5. https://doi.org/10.1007%2Fs41116-017-0009-6

Li, Y., Luhmann, J. G., Lynch, B. J.: 2018, Magnetic clouds: solar cycle dependence, sources, and geomagnetic impacts. *Solar Phys*. **293**(10), 135. https://doi.org/10.1007%2Fs11207-018-1356-8

Lefèvre, L., Vennerstrøm, S., Dumbović, M, Vršnak, B., Sudar, D., Arlt, R., *et al*.: 2016, Detailed analysis of solar data related to historical extreme geomagnetic storms: 1868 – 2010. *Solar Phys*. **291**(5), 1483–2016. https://doi.org/10.1007/s11207-016-0892-3

Liu, Y. D., Luhmann, J. G., Kajdič, P., Kilpua, E. K. J., Lugaz, N., Nitta, N. V., *et al*.: 2014, Observations of an extreme storm in interplanetary space caused by successive coronal mass ejections. *Nature Communications*. **5**, 3481. https://doi.org/10.1038/ncomms4481

Love, J. J., Hayakawa, H., Cliver, E. W.: 2019, On the intensity of the magnetic superstorm of September 1909. *Space Weather*, **17**(1), 37–45. https://doi.org/10.1029/2018SW002079

Manchester, W. IV, Kilpua, E. K. J., Liu, Y. D., Lugaz, N., Riley, P., Török, T., Vršnak, B.: 2017. The physical processes of CME/ICME evolution. *Space Sci. Rev.* **212**(3–4), 1159–1219. https://doi.org/10.1007/s11214-017-0394-0

McAllister, A. H., Dryer, M., Mcintosh, P., Singer, H., Weiss, L.: 1996, A large polar crown CME and a "problem" geomagnetic storm: April 14-23, 1994. *J. Geophys. Res*. **101**(A6), 13,497–13,515. https://doi.org/10.1029/96JA00510

Michalek, G., Gopalswamy, N., Yashiro, S.: 2008, Space weather application using projected velocity asymmetry of halo CMEs. *Solar Phys*. **248**(1), 113–123. https://doi.org/10.1007/s11207-008-9126-7

Newton, H. W.: 1943, Solar flares and magnetic storms. *Mon. Not. R. Astron. Soc.* **103**(5), 244–257. https://doi.org/doi:10.1093/mnras/103.5.244

Newton, H. W.: 1944, Solar flares and magnetic storms (second paper), *Mon. Not. R. Astron. Soc.* **104**(1), 4–12. https://doi.org/10.1093/mnras/104.1.4

Pal, S., Nandy, D., Srivastava, N., Gopalswamy, N., Panda, S.: 2018, Dependence of coronal mass ejection properties on their solar source active region characteristics and associated flare reconnection flux. *Astrophys. J.* **865**(1), 4. https://doi.org/10.3847/1538-4357/aada10

Richardson, I. G., Cane, H.V.: 2010, Near-Earth interplanetary coronal mass ejections during solar cycle 23 (1996 – 2009): catalog and summary of properties. *Solar Phys*. **264**(1), 189–237.[1] https://doi.org/10.1007/s11207-010-9568-6

Shanmugaraju, A., Ibrahim, M. S., Moon, Y.-J., Rahman, A. M., Umapathy, S.: 2015, Empirical relationship between CME parameters and geo-effectiveness of halo CMEs in the rising phase of solar cycle 24 (2011–2013). *Solar Phys*. 290(5), 1417–1427. https://doi.org/10.1007/s11207-015-0671-6

Srivastava, N., Venkatakrishnan, P.: 2004, Solar and interplanetary sources of major geomagnetic storms during 1996–2002. *J. Geophys. Res*. **109**(A10), 10,103. https://doi.org/10.1029/2003JA010175

Sugiura, M., Kamei, T.: 1991, Equatorial Dst index 1957–1986. *IAGA Bulletin*, **40**, 1–246.[2]

Tsurutani, B., Gonzalez, W. D., Tang, F., Lee, Y. T.: 1992, Great magnetic storms. *Geophys. Res. Lett.* **19**(1), 73–76. https://doi.org/10.1029/91GL02783

Vaisberg, O.L., Zastenker, G.N.: 1976, Solar wind and magnetosheath observations at Earth during August 1972. *Space Sci. Rev*. **19**(4–5), 687–702. https://doi.org//10.1007%2FBF00210646

Wang, R., Liu, Y. D., Hu, H., Zhao, X.: 2018, A solar eruption with relatively strong geo-effectiveness originating from active region peripheral diffusive polarities. *Astrophys. J.* **863**(1), 81. https://doi.org/10.3847/1538-4357/aad22d

Webb, D., Howard, T. A.: 2012, Coronal mass ejections: observations. *Living Rev. Sol. Phys.* **9**, 3. https://doi.org/10.12942/lrsp-2012-3

Webb, D., Nitta, N.: 2017, Understanding problem forecasts of ISEST campaign flare-CME events. *Solar Phys*. **292**(10), 142. https://doi.org/10.1007/s11207-017-1166-4

Wu, C-C., Liou, Ronald, K., Lepping, P., Hutting, L.: 2019, The 04–10 September 2017 Sun–Earth connection events: solar flares, coronal mass ejections/magnetic clouds, and geomagnetic storms. *Solar Phys*. **294**(8), 110. https://doi.org/10.1007/s11207-019-1446-2

Yashiro, S., Gopalswamy, N., Michalek, G., St.Cyr, O. C., Plunkett, S. P., Rich, N. B., Howard, R. A.: 2004, A catalog of white light coronal mass ejections observed by the SOHO spacecraft. *J. Geophys. Res*. **109**(A7), A07105[3]. https://doi.org/10.1029/2003JA010282

Yermolaev, Yu. I., Lodkina, I. G., Nikolaeva, N. S., Yermolaev, M. Yu.: 2018, Geoeffectiveness of Solar and Interplanetary Structures and Generation of Strong Geomagnetic Storms. In: Extreme Events in Geospace. Origins, predictability, and consequences. Ed. by N. Buzulukova, Elsevier, p. 99–113. https://doi.org/10.1016/B978-0-12-812700-1.00004-2

Zhang, J., Richardson, I. G., Webb, D. F., Gopalswamy, N., Huttunen, E., Kasper, J., *et al*.: 2007a, Solar and interplanetary sources of major geomagnetic storms (Dst ≤100 nT) during 1996–2005. *J. Geophys. Res*. **112**(A10)**,** 10102**.** https://doi.org/doi:10.1029/2007JA012321

Zhang, J., Richardson, I. G., Webb, D. F., Gopalswamy, N., Huttunen, E., Kasper, J., *et al*.: 2007b, Correction to "Solar and interplanetary sources of major geomagnetic storms (Dst ≤100 nT) during 1996–2005", *J. Geophys. Res*. **112**(A12), 12103. https://doi.org/10.1029/2007JA012891

Zou, P., Jiang, C., Wei, F., Zuo, P., Wang, Y.: 2019, A statistical study of solar filament eruptions that forms high-speed coronal mass ejections. *Astrophys. J.* **884**(2),157. https://doi.org/10.3847/1538-4357/ab4355

---

[1] See data after 2009 at http://www.srl.caltech.edu/ACE/ASC/DATA/level3/icmetable2.htm
[2] See http://wdc.kugi.kyoto-u.ac.jp/Dstdir/Dst2/onDstindex.html
[3] See https://cdaw.gsfc.nasa.gov/CME_list/